\documentclass[aps,preprint]{revtex4}
\usepackage{amssymb}
\usepackage{amsfonts}
\usepackage{amsmath}
\usepackage{graphicx}

\begin{document}

\title{Conservation law of energy-momentum in general relativity}
\author{Yi-Shi Duan and Jing-Ye Zhang}
\affiliation{Lanzhou University}

\begin{abstract}
We  explain the necessity of application of semi-metric in general relativity. A detailed discussion on the energy-momentum conservation in the general
relativity is presented using the mathematical tool of semi-metric. By means of the general covariant spacetime translation transformation,
the most general covariant conservation law of energy-momentum is obtained,
which is valid for any coordinates and overcomes the flaws of the expressions of
Einstain, Landau and Moller.
\end{abstract}

%\date[Nov. 1963]{\today}
\received[Received on ]{Oct. 31, 1962}

\published[Published in ]{Nov. 1963}

\maketitle

\section{Introduction}

The energy-momentum conservation in the general relativity still remains an open
question. Since there is no reasonable expression for gravitational energy as of today,
the radiating of gravitational field and other important issues in
general relativity can't be solved. Therefore, the energy-momentum
conservation law is one of the most fundamental questions in general
relativity. There are some discussions about this topic [1] recently.
However, there is no satisfied answer yet.

In the early stage of general relativity, Einstein and Tolman proposed a
general conservation law as [2,3]%
\begin{equation}
\frac{\partial\sqrt{g}\theta_{i}^{k}}{\partial x^{k}}=0,
\end{equation}%
\begin{equation}
\theta_{i}^{k}=T_{i}^{k}+t_{i,can}^{k},
\end{equation}
where $T_{i}^{k}$ is the energy-momentum tensor for matter appearing on the
r.h.s of the Einstein equation%
\begin{equation}
R_{i}^{k}-\frac{1}{2}\delta_{i}^{k}R=\frac{8\pi k}{c^{4}}T_{i}^{k}
\end{equation}
and $t_{i,can}^{k}$ is the canonical energy-momentum tensor of gravitational
field which is a pseudotensor or affine tensor, instead of a Riemann tensor,
i.e, it can be regarded as a tensor only with the linear transformation, %
\begin{equation}
t_{i,can}^{k}=\frac{1}{\sqrt{g}}\left[ \delta_{i}^{k}L_{g}-\frac{\partial
L_{g}}{\partial g_{k}^{lm}}g_{i}^{lm}\right] .
\end{equation}
$L_{g}$ is the Lagangian of gravitional field $L_{g}$ multiply with $\sqrt{g}
$, i.e.%
\begin{equation*}
L_{g}=\sqrt{g}L_{g},
\end{equation*}%
\begin{equation}
L_{g}=\frac{c^{4}}{16\pi k}g^{kl}\left[ \Gamma_{lj}^{i}\Gamma_{ki}^{j}-%
\Gamma_{lk}^{i}\Gamma_{ij}^{j}\right] .
\end{equation}
With Eq. (1) and the four dimensional Gaussian theorem, one obtains the
four-momentum for the closed system%
\begin{equation}
P_{i}=\frac{1}{ic}\int\theta_{i}^{k}\sqrt{g}d\sigma_{k},\ \ \ (i=1,2,3,4)
\end{equation}
or%
\begin{equation}
P_{i}=\frac{1}{ic}\int\theta_{i}^{4}\sqrt{g}dV
\end{equation}
which is a time-independent conserved quantity.

H. Bauer [4] and E. Schrondinger [5] realized that the above definition of
energy-momentum tensor has a serious problem. Based on the definition (2) and
(6), the total energy of a closed system is finite and reasonable only in the
quasi-Galilean coordinates. In other coordinates, for example the spherical
coordinate, such total energy is divergent. In addition, the energy density
of vacuum (without any matter and gravitational field) should be zero.
However, such energy density in the spherical coordinates is nonzero with
their definition.

Landau [6] once adopted a different approach and suggested that the energy density of gravitational field should
be described by the symmetric energy-momentum pesudotensor $t^{ik}$ and
the four momentum should carry the contravariant index in order to obtain the
conservation law of moment. However, his theory is also only valid in the
quasi-Galilean coordinates and suffers the same flaws as Einstein's proposal.

C.Moller [7] investigated recently the problem of energy conservation law in
general relativity and found that the essential reason for Einstain's
expression of gravitational energy not working in the non-quasi-Galilean
coordinate is that the $\theta_{i}^{4}$ in Eq. (7) is not a vector under the
pure space coordinate transformation. In order to solve this problem, Moller
proposed a new expression for energy-momentum tensor which gives the total
four momentum in a closed system independent of the choice of coordinates.
Furthermore he argued that new theory is valid even for an non-closed system.
However,  Moller himself [8] later realized in 1960 that the new $\theta_{i}^{4}$
includes a term decay to zero as $1/r^{3}$ when $r$ goes to infinity.
Therefore, this theory still can not specify the correct energy-momentum conservation
law.

From Eq. (6) one can find that index $k$ of $\theta_{i}^{k}$ or $t_{i}^{k}$
will be summed with $\sqrt{g}d\sigma_{k}$ and other index $i$ give the index
for momentum $P_{i}$. Since $P_{i}$ is a conserved quantity for a closed
system which can undergo only inertial motions, and initial frames are connected by the Lorentz
transformations, the index $i$ of $\theta_{i}^{k}$ corresponding to momentum $%
P_{i}$ shows the property of tensor only under Lorentz transformation, i.e. $%
\theta_{i}^{k}$ is an affine tensor for index $i$. Therefore, it is
reasonable that the index $i$ only allows the Lorentz transformation.
However, both indices $k,i$ of the energy-momentum tensor $\theta_{i}^{k}$ defined by Einstein and
Tolman are affine indices instead of Riemann indices, i.e., except index $i$, the
index $k$ sum with volume element $\sqrt{g}d\sigma_{k}$ is a tensor index
only for linear transformation (the same as the Landau's theory).
It is well recognized nowadays that this is the main reason for those theories give total
energy dependent on the choice of coordinates. Moller and others are paying efforts recently to improve the covariance of $\theta_{i}^{k}$.
We believe, this problem can be solved if the index $k$ in $\theta_{i}^{k}$
is a Riemann index hence the sum with $\sqrt{g}d\sigma_{k}$ is a Riemann tensor contraction.
At the same time, with the momentum index $i$ of $P_{i}$  given a special meaning in a certain sense, the flaws in the old expressions can be overcome.

In addition, the physical reason that the previous definition of energy
momentum tensor is valid only in the quasi-Galilean coordinates is that
those energy-momentum tensors include the inertial force field besides
gravitational field.

In principle, any physical law including energy-momentum conservation
law in general relativity should be generally covariant (except the coordinate
condition). Theories satisfying above criteria will be valid in any
coordinates. Furthermore, the four momentum will not contradict with
the equivalent principle if the momentum includes only gravitational field while excluding
inertial force field. The choice of coordinate condition is to fix the
reference coordinates, therefore the coordinate condition should not be
generally covariant (or else $g_{ik}$ will be the same for all coordinates).
It seems impossible to obtain generally covariant expression for
energy-momentum tensor in the general relativity because the conservation
law (1) is not generally covariant even if the energy-momentum tensor $%
\theta_{i}^{k}$ is a Riemann tensor. The reason is that we can't obtain the
conserved quantity (6) with Gauss theory if eq.(1) is written in a generally covariant form,%
\begin{equation*}
\left( \theta_{i}^{k}\right) _{k}=\frac{1}{\sqrt{g}}\frac{\partial\left(
\sqrt{g}\theta_{i}^{k}\right) }{\partial x^{k}}-\frac{1}{2}\frac{\partial
g_{kl}}{\partial x^{i}}\theta^{kl}=0.
\end{equation*}
We believe that afore mentioned problems can be solved by introducing the semi-metric.
Moller mentioned similar idea in his recent work [8], but the covariant
conservation law and unique problem have not been solved so far.

In this paper, we  explain the necessity of application of semi-metric in general relativity. A detailed discussion on the energy-momentum conservation in the general
relativity is presented using the mathematical tool of semi-metric. By means of the general covariant spacetime translation transformation,
the most general covariant conservation law of energy-momentum is obtained,
which is valid for any coordinates and overcomes the flaws of the expressions of
Einstain, Landau and Moller.

\section{Semi-metric}

Gravitational field can be described by metric $g_{ik}$. However, It can be described by semi-metric just as well. We will mainly discuss this tool in this
section.

The relation between metric and semi-metric is [9,11,12,20]%
\begin{equation*}
g_{ik}=\lambda_{i\left( \alpha\right) }\lambda_{k\left( \alpha\right) },\ \
\ \ \ \left\Vert g_{ik}\right\Vert =\left\Vert \lambda_{i\left(
\alpha\right) }\right\Vert ^{2},
\end{equation*}%
\begin{equation}
\sqrt{g}=\sqrt{|g_{ik}|}=|\lambda_{i\left( \alpha\right) }|\neq0;
\end{equation}
and%
\begin{equation*}
\lambda_{\left( \alpha\right) }^{i}\lambda_{i\left( \beta\right)
}=\delta_{\alpha\beta},
\end{equation*}%
\begin{equation}
\lambda_{\left( \alpha\right) }^{i}\lambda_{j\left( \alpha\right)
}=\delta_{j}^{i};
\end{equation}
where $\lambda_{\left( \alpha\right) }^{i}$ is the element of inverse matrix
$\left\Vert \lambda_{i\left( \alpha\right) }\right\Vert ^{-1}$, $%
\lambda_{i\left( \alpha\right) }$ and $\lambda_{\left( \alpha\right) }^{i}$
are  Riemann tensors for index $i$. Later we call
$g_{ik}$ as the fundamental field metric representation while
$\lambda_{\left( \alpha\right) }^{i}$ as the fundamental field semi-metric representation. (The
unique problem of $\lambda_{i\left( \alpha\right) }$ will be discussed in
the appredix IV.)

With (8), the orthogonal transformation group of $\lambda_{i\left(
\alpha\right) }$ keeps $g_{ik}$ invariant%
\begin{equation*}
\lambda_{i\left( \alpha\right) }^{\prime}=L_{\left( \alpha\beta\right)
}\lambda_{i\left( \beta\right) }
\end{equation*}%
\begin{equation}
L_{\left( \alpha\beta\right) }L_{\left( \alpha\gamma\right) }=\delta_{\left(
\beta\gamma\right) },
\end{equation}
i.e.%
\begin{equation}
g_{ik}^{\prime}=\lambda_{i\left( \alpha\right) }^{\prime}\lambda_{k\left(
\alpha\right) }^{\prime}=\lambda_{i\left( \alpha\right) }\lambda_{k\left(
\alpha\right) }=g_{ik}.
\end{equation}
With the definition%
\begin{equation*}
dx_{\left( \alpha\right) }\equiv\lambda_{i\left( \alpha\right) }dx^{i}
\end{equation*}
or%
\begin{equation}
dx^{i}=\lambda_{\left( \alpha\right) }^{i}dx_{\left( \alpha\right) },\ \ \ \
x^{4}=ict,
\end{equation}
one can obtain
\begin{equation}
dS^{2}=-g_{ik}dx^{i}dx^{k}=-dx_{\left( \alpha\right) }dx_{\left(
\alpha\right) }.
\end{equation}
From above result one can find that $dS^{2}$ is an invariant for the
orthogonal transformation $\left( 10\right) $ or
\begin{equation}
dx_{\left( \alpha\right) }^{\prime}=L_{\left( \alpha\beta\right) }dx_{\left(
\beta\right) }\text{.}
\end{equation}
However, in general $dx_{\left( \alpha\right) }$ is not a total derivative.
Only in the region where the gravitational field vanishes can it be regarded
as a total derivative. In this case $dS$ defined in (13) is that in the
psedo-Euclidean space. Therefore, in the gravity-free region, orthogonal
transformation group (10) or (14) is just the usual Lorentz group.

It is seen from above discussion that there are two kinds of transformation groups in the
general relativity

(a) general transformation group of Riemann tensor%
\begin{equation}
x^{i\prime}=f^{i}\left( x^{1},x^{2},x^{3},x^{4}\right) \left\vert \frac{%
\partial f^{i}}{\partial x^{k}}\right\vert \neq0;
\end{equation}

(b) orthogonal tranformation group of semi-metric tensor%
\begin{equation*}
\lambda _{i\left( \alpha \right) }^{\prime }=L_{\left( \alpha \beta \right)
}\lambda _{i\left( \beta \right) }
\end{equation*}%
or%
\begin{equation}
dx_{\left( \alpha \right) }^{\prime }=L_{\left( \alpha \beta \right)
}dx_{\left( \beta \right) },\ \ \ \ \left\vert L_{\left( \alpha \beta
\right) }\right\vert \neq 0.
\end{equation}%
Every physical law should keep covariant respect to the above two transformation groups.

Next let's construct the theory of general relativity from the viewpoint of
semi-metric $\lambda_{i\left( a\right) }$. Mathematically there is no
essential difference to work with either metric representation or
semi-metric representation. However, the semi-metric as the
fundamental field of gravity has more general meaning than the metric in physics. For
example, we can write the reasonable interaction between gravity and other
particles, especially with the spinor fields, only in the semi-metric
representation instead of metric one. Moller, Fock, Infeld [8,20,16] and
one of authors [14,15] all pointed out this problem. In this work we will
show that the energy-momentum conservation law in the general relativity can
only be solved with semi-metric.

\bigskip In order to present the theory of general relativity based on
semi-metric, we need derive the Einstein equations (3) by varying $\lambda
_{i(a)}$ based on the principle of least action. Let $I$ is the action of
gravity and matter%
\begin{equation}
I=\int L\sqrt{g}\left( dx\right) =\int L\left( dx\right) ,
\end{equation}%
where%
\begin{equation*}
L=L_{\lambda }+L_{m},\ \ \ \ \ L\equiv \sqrt{g}L;
\end{equation*}%
$L_{\lambda }$ is the Larangian of gravity in the semi-metric representation,
$L_{m}$ is the Lagrangian for matter and $L$ is the total Lagangian.
According to the requirement of general covariance, the action $I$ should be
Riemannian scalar. Since $\sqrt{g}\left( dx\right) $ is a scalar volume
element, $L_{\lambda }$ and $L_{m}$ should also be the Riemannian scalars.
In order to fulfill the requirement, we follow the method of ordinary metric
to find $L_{\lambda }$ from Riemannian scalar curvature $R$.

There is the relation between curvation $R_{jlk}^{m}$ and semi-metric $%
\lambda _{\left( a\right) }^{m}$%
\begin{equation}
\left( \lambda _{\left( a\right) }^{m}\right) _{ik}-\left( \lambda _{\left(
a\right) }^{m}\right) _{kj}=-\lambda _{\left( a\right) }^{l}R_{ljk}^{m}.
\end{equation}%
With%
\begin{equation}
R_{lk}=R_{lmk}^{m},\ \ \ R=g^{lk}R_{lk}
\end{equation}%
and Eq. (9) we can find
\begin{equation}
R=\lambda _{\left( a\right) }^{k}\left[ \left( \lambda _{\left( a\right)
}^{j}\right) _{kj}-\left( \lambda _{\left( a\right) }^{j}\right) _{jk}\right]
;
\end{equation}%
With $\left( g^{kj}\right) _{l}=0$ one obtain%
\begin{equation}
\lambda _{\left( a\right) }^{k}\left( \lambda _{\left( a\right) }^{j}\right)
_{l}=-\lambda _{\left( a\right) }^{j}\left( \lambda _{\left( a\right)
}^{k}\right) _{l}.
\end{equation}%
It is easy to find%
\begin{equation*}
\sqrt{g}R=\sqrt{g}G-2\frac{\partial }{\partial x^{i}}\left[ \sqrt{g}\lambda
_{\left( a\right) }^{i}\left( \lambda _{\left( a\right) }^{j}\right) _{j}%
\right] ,
\end{equation*}%
and
\begin{equation}
G\equiv \left[ \left( \lambda _{\left( a\right) }^{i}\right) _{i}\left(
\lambda _{\left( a\right) }^{j}\right) _{j}-\left( \lambda _{\left( a\right)
}^{i}\right) _{j}\left( \lambda _{\left( a\right) }^{j}\right) _{i}\right] .
\end{equation}%
The second term of r.h.s for $\sqrt{g}R$ is a 4-dimensional divergence. With
Gauss theorem and the variation should be zero at the boundary, one can prove that%
\begin{equation}
\delta \int R\sqrt{g}\left( dx\right) =\delta \int G\sqrt{g}\left( dx\right)
.
\end{equation}%
Therefore we can define the Lagangian with semi-metric as $\frac{c^{4}}{16\pi
k}G$, i.e.%
\begin{equation}
L_{\lambda }=\frac{c^{4}}{16\pi k}\left[ \left( \lambda _{\left( a\right)
}^{i}\right) _{i}\left( \lambda _{\left( a\right) }^{j}\right) _{j}-\left(
\lambda _{\left( a\right) }^{i}\right) _{j}\left( \lambda _{\left( a\right)
}^{j}\right) _{i}\right] .
\end{equation}%
With the relation between covariant derivative of semi-metric and Ricci
coefficient $\eta _{\left( \alpha \beta \gamma \right) }$%
\begin{equation}
\left( \lambda _{\left( \beta \right) }^{i}\right) _{j}=\lambda _{\left(
\gamma \right) }^{i}\lambda _{j\left( \alpha \right) }\eta _{\left( \alpha
\beta \gamma \right) }
\end{equation}%
$L_{\lambda }$ can be expressed as%
\begin{equation}
L_{\lambda }=\frac{c^{4}}{16\pi k}\left[ \eta _{\left( \alpha \right) }\eta
_{\left( \alpha \right) }-\eta _{\left( \alpha \beta \gamma \right) }\eta
_{\left( \gamma \beta \alpha \right) }\right] ,
\end{equation}%
where%
\begin{eqnarray}
\eta _{\left( \alpha \beta \gamma \right) } &=&\frac{1}{2}\{\frac{\partial
\lambda _{\sigma \left( \alpha \right) }}{\partial x^{\lambda }}\left[
\lambda _{\left( \beta \right) }^{\lambda }\lambda _{\left( \gamma \right)
}^{\sigma }-\lambda _{\left( \gamma \right) }^{\lambda }\lambda _{\left(
\beta \right) }^{\sigma }\right]  \notag \\
&&+\frac{\partial \lambda _{\sigma \left( \beta \right) }}{\partial
x^{\lambda }}\left[ \lambda _{\left( \alpha \right) }^{\lambda }\lambda
_{\left( \gamma \right) }^{\sigma }-\lambda _{\left( \gamma \right)
}^{\lambda }\lambda _{\left( \alpha \right) }^{\sigma }\right]  \notag \\
&&+\frac{\partial \lambda _{\sigma \left( \gamma \right) }}{\partial
x^{\lambda }}\left[ \lambda _{\left( \beta \right) }^{\lambda }\lambda
_{\left( \alpha \right) }^{\sigma }-\lambda _{\left( \alpha \right)
}^{\lambda }\lambda _{\left( \beta \right) }^{\sigma }\right] ,
\end{eqnarray}%
\begin{equation}
\eta _{\left( \alpha \right) }=\eta _{\left( \beta \alpha \beta \right) }=%
\frac{1}{\sqrt{g}}\frac{\partial \sqrt{g}\lambda _{\left( \alpha \right)
}^{i}}{\partial x^{i}}=\left( \lambda _{\left( \alpha \right) }^{i}\right)
_{i}.
\end{equation}%
There is an important difference between $L_{\lambda }$ and the Lagrangian $%
L_{g}$ in Eq. (5) with metric. Firstly,  one can find with (24) that $%
L_{\lambda }$ is a Riemannian scalar expressed by covariant derivative of $%
\lambda _{\left( \alpha \right) }^{i}$, which is essential to obtaining the
general covariant conservation law. In contrast $L_{g}$ defined in (5) is
not a Riemannian scalar hence lacking above properties. Secondly, it is easy to
show that the difference between $\sqrt{g}L_{\lambda }$ and $\sqrt{g}L_{g}$
is just a 4-dimensional divengence $\Delta $:%
\begin{equation*}
\sqrt{g}L_{\lambda }=\sqrt{g}L_{g}+\Delta ,
\end{equation*}%
\begin{equation}
\Delta =\frac{c^{4}}{16\pi k}\frac{\partial }{\partial x^{i}}\left[ \sqrt{g}%
\left( \lambda _{\left( \alpha \right) }^{i}\frac{\partial \lambda _{\left(
\alpha \right) }^{j}}{\partial x^{j}}-\lambda _{\left( \alpha \right) }^{j}%
\frac{\partial \lambda _{\left( \alpha \right) }^{i}}{\partial x^{j}}\right) %
\right] .
\end{equation}%
Later we will find that this divengent term is useful to get the
conservation law, although it will not contribute to the variation of the
action and equations of motion.

Now let's derive the equations of gravity by the principle of least action
\begin{equation}
\delta I=\delta \int L\left( dx\right) =0.
\end{equation}%
With%
\begin{equation}
\delta \int L\left( dx\right) =\int \left[ L\right] _{\lambda _{\left(
\alpha \right) }^{i}}\delta \lambda _{\left( \alpha \right) }^{i}\left(
dx\right) ,
\end{equation}%
\begin{equation}
\left[ L\right] _{\lambda _{\left( \alpha \right) }^{i}}\equiv \frac{%
\partial L}{\partial \lambda _{\left( \alpha \right) }^{i}}-\frac{\partial }{%
\partial x^{j}}\frac{\partial L}{\partial \lambda _{\left( \alpha \right)
j}^{i}},\ \ \ \ \lambda _{\left( \alpha \right) j}^{i}\equiv \frac{\partial
\lambda _{\left( \alpha \right) }^{i}}{\partial x^{j}}.
\end{equation}%
and (30) one finds%
\begin{equation*}
\left[ L\right] _{\lambda _{\left( \alpha \right) }^{i}}=0,
\end{equation*}%
i.e.%
\begin{equation}
\left[ L_{\lambda }\right] _{\lambda _{\left( \alpha \right) }^{i}}+\left[
L_{m}\right] _{\lambda _{\left( \alpha \right) }^{i}}=0,
\end{equation}%
where%
\begin{equation*}
L_{\lambda }\equiv \sqrt{g}L_{\lambda },\ \ \ L_{m}\equiv \sqrt{g}L_{m}.
\end{equation*}%
On the other hand, with%
\begin{equation}
\delta \int G\sqrt{g}\left( dx\right) =\frac{16\pi k}{c^{4}}\delta \int
L_{\lambda }\left( dx\right) =\frac{16\pi k}{c^{4}}\int \left[ L_{\lambda }%
\right] _{\lambda _{\left( \alpha \right) }^{i}}\delta \lambda _{\left(
\alpha \right) }^{i}\left( dx\right) ,
\end{equation}%
and the ref. [6], Eq. (8) and the symmetry of $g_{ik}$ and $R_{ik}$%
\begin{eqnarray}
\delta \int R\sqrt{g}\left( dx\right) &=&\int \left( R_{ik}-\frac{1}{2}%
g_{ik}R\right) \delta g^{ik}\sqrt{g}\left( dx\right)  \notag \\
&=&\int \left( R_{ik}-\frac{1}{2}g_{ik}R\right) \left( \lambda _{\left(
\alpha \right) }^{i}\delta \lambda _{\left( \alpha \right) }^{k}+\lambda
_{\left( \alpha \right) }^{k}\delta \lambda _{\left( \alpha \right)
}^{i}\right) \sqrt{g}\left( dx\right)  \notag \\
&=&2\int \left( R_{ik}-\frac{1}{2}g_{ik}R\right) \lambda _{\left( \alpha
\right) }^{k}\delta \lambda _{\left( \alpha \right) }^{i}\sqrt{g}\left(
dx\right) ,
\end{eqnarray}%
we can find from (23)%
\begin{equation}
R_{ik}-\frac{1}{2}g_{ik}R=\frac{8\pi k}{c^{4}}\frac{1}{\sqrt{g}}\left[
L_{\lambda }\right] _{\lambda _{\left( \alpha \right) }^{i}}\lambda
_{k\left( \alpha \right) }.
\end{equation}%
Defining the energy-momentum tensor of matter as%
\begin{equation}
T_{i\left( \alpha \right) }\equiv -\frac{1}{\sqrt{g}}\left[ L_{m}\right]
_{\lambda _{\left( \alpha \right) }^{i}},
\end{equation}%
\begin{equation*}
T_{ik}=T_{i\left( \alpha \right) }\lambda _{k\left( \alpha \right) };
\end{equation*}%
with (33) we can immediately get the Einstein equations%
\begin{equation}
R_{ik}-\frac{1}{2}g_{ik}R=\frac{8\pi k}{c^{4}}T_{ik}.
\end{equation}

Then we show that the Einstein equations could be obtained by varing $%
\lambda _{\left( \alpha \right) }^{i}$ with the gravitaional Lagrangian $L_{\lambda }$
defined in the semi-metric representation (24). This clearly proves that
semi-metric $\lambda _{\left( \alpha \right) }^{i}$ can be regarded in deed as the
fundamental field of gravity.

\section{Generally covariant energy-moentum conservation law}

In this section we will study the generally covariant conservation law based
on the Lagrangian defined by semi-metric in previous section. What we are about to obtain is a
conservation law that is generally covariant, i.e. the conservation law
valids in the arbitrary coordinates.  This will solve the long-standing
problem of energy-momentum conservation law pointed out by Einstain, Landau
and Moller discussed in the introduction.

In the classical field theory, Noether theorem tells us that the invariance
of the total action of a system under certain transformations correspond to
a conserved quantities. We will study the energy-momentum conservation law
in general relativity based on such a well established viewpoint.

The action
\begin{equation}
I=\int_{G}L\left( v^{A},v_{i}^{A}\right) \left( dx\right) =\int_{G}L\left(
v^{A},v_{i}^{A}\right) \sqrt{g}\left( dx\right)
\end{equation}%
keeps invariant under the infinitesimal transformation%
\begin{equation*}
x\rightarrow x^{\prime }=x+\delta x
\end{equation*}%
\begin{equation}
v^{A}\left( x\right) \rightarrow v^{A}\left( x^{\prime }\right) ^{\prime
}=v^{A}\left( x\right) +\delta v^{A}\left( x\right) .
\end{equation}%
where $v^{A}\left( x\right) $ are any fields including gravitational field
and matter field, $A$ is the index for the component of fields, $%
v_{i}^{A}\left( x\right) \equiv \frac{\partial v^{A}\left( x\right) }{%
\partial x^{i}}$. Suppose $\delta v^{A}\left( x\right) $ vanishes at the
boundary of 4-dimensional volume $G$, then one can prove that (see Ref. [17]
and [18]):%
\begin{equation}
\frac{\partial }{\partial x^{k}}\left( L\delta x^{k}-\frac{\partial L}{%
\partial v_{k}^{A}}v_{l}^{A}\delta x^{l}+\frac{\partial L}{\partial v_{k}^{A}%
}\delta v^{A}\right) +\left[ L\right] _{v^{A}}\left( \delta
v^{A}-v_{l}^{A}\delta x^{l}\right) =0.
\end{equation}%
If $L$ is the total action of the system, with the principle of least action
$\delta I=0$, varying $L$ by $v^{A}$ gives the Euler equation%
\begin{equation}
\left[ L\right] _{v^{A}}=\frac{\partial L}{\partial v^{A}}-\frac{\partial }{%
\partial x^{i}}\frac{\partial L}{\partial v_{i}^{A}}=0.
\end{equation}%
Then with (41), for transformation (40) there is a conservation law%
\begin{equation}
\frac{\partial }{\partial x^{k}}\left( L\delta x^{k}-\frac{\partial L}{%
\partial v_{k}^{A}}v_{l}^{A}\delta x^{l}+\frac{\partial L}{\partial x_{k}^{A}%
}\delta v^{A}\right) =0.
\end{equation}%
Eq. (42) is the equation of motion for $v^{A}$ from the principle of least
action.

Here we want to emphasis that (41) is valid only if $L$ is invariant for
transformation (40), even $L$ in (41) is not the total Langangian. However,
in case $I$ is not the total action, $\delta I\neq 0$ leads (42) and (43)
invalid.

When we study the gravitational field with semi-metric representation, the
Lagrangian in (39) should include the Riemannian scalar with the semi-metric $%
\lambda _{\left( \alpha \right) }^{i}$ and its first order derivative. Since
$\lambda _{\left( \alpha \right) }^{i}$ is a contravariant tensor, the
action $I$ is invariant under the following transformation%
\begin{equation}
x^{\prime }=x+\delta x,\ \ \ \ \ \lambda _{\left( \alpha \right) }^{i\prime
}=\frac{\partial x^{i\prime }}{\partial x^{l}}\lambda _{\left( \alpha
\right) }^{l}.
\end{equation}%
The second transformation in (44) can be written with general infintesimal
transformation%
\begin{equation}
\lambda _{\left( \alpha \right) }^{i\prime }=\lambda _{\left( \alpha \right)
}^{i}+\delta \lambda _{\left( \alpha \right) }^{i},\ \ \ \ \delta \lambda
_{\left( \alpha \right) }^{i}=\frac{\partial \delta x^{i}}{\partial x^{l}}%
\lambda _{\left( \alpha \right) }^{l}.
\end{equation}%
When the field in (41)is the gravitational field $\lambda _{\left( \alpha
\right) }^{i}$, i.e. $v^{A}=\lambda _{\left( \alpha \right) }^{i}$, one can
find an important result%
\begin{eqnarray}
&&\frac{\partial }{\partial x^{k}}[\left( L\delta _{l}^{k}-\frac{\partial L}{%
\partial \lambda _{\left( \alpha \right) k}^{i}}\lambda _{\left( \alpha
\right) l}^{i}\right) \delta x^{l}+\frac{\partial L}{\partial \lambda
_{\left( \alpha \right) k}^{i}}\lambda _{\left( \alpha \right) }^{l}\frac{%
\partial \delta x^{i}}{\partial x^{l}}]  \notag \\
&&+\left[ L\right] _{\lambda _{\left( \alpha \right) }^{i}}\left[ \frac{%
\partial \delta x^{i}}{\partial x^{l}}\lambda _{\left( \alpha \right)
}^{l}-\lambda _{\left( \alpha \right) l}^{i}\delta x^{l}\right]
=0.
\end{eqnarray}%
When $I$ is the total action, i.e. $L=L_{\lambda }+L_{m}$, we can get the
equations for gravitational field corresponding to (42)%
\begin{equation}
\left[ L\right] _{\lambda _{\left( \alpha \right) }^{i}}=0,
\end{equation}%
and the conservation law corresponding to (43)%
\begin{equation}
\frac{\partial }{\partial x^{k}}\left[ \left( L\delta _{l}^{k}-\frac{%
\partial L}{\partial \lambda _{\left( \alpha \right) k}^{i}}\lambda _{\left(
\alpha \right) l}^{i}\right) \delta x^{l}+\frac{\partial L}{\partial \lambda
_{\left( \alpha \right) k}^{i}}\lambda _{\left( \alpha \right) }^{l}\frac{%
\partial \delta x^{i}}{\partial x^{l}}\right] =0.
\end{equation}%
From previous discussion one can find that (47) is just the Einstein
equtions for gravity. The conserved quantities determined by (48) and Gauss theorem is
specified by total Lagrangian $L$.\ An important property of gravitational
theory is that the conservation law of gravitational field and matter field
can be expressed by the Lagrangian of gravitational field $L_{\lambda }$
only, independent of the matter field $L_{m}$. As will be discussed later on, this unique feature is directly related
to some properties of Einstein equation. In the following, one can find that it is very
convenient to study the concrete problems when the conservation law is
expressed only by the gravitational field $L_{\lambda }$.

In order to express the conservation law only by the gravitational field, we
replace the total $L$ by the $L_{\lambda }$ in Eq. (46). Since $L_{\lambda }$
is invariant under the transformation (44) and (45), Eq. (46) is still
correct. However, since $\left[ L\right] _{\lambda _{\left( \alpha \right)
}^{i}}$ is not zero, (46) can be expressed as%
\begin{eqnarray}
&&\frac{\partial }{\partial x^{k}}[\left( L_{\lambda }\delta _{l}^{k}-\frac{%
\partial L_{\lambda }}{\partial \lambda _{\left( \alpha \right) k}^{i}}%
\lambda _{\left( \alpha \right) l}^{i}\right) \delta x^{l}+\frac{\partial
L_{\lambda }}{\partial \lambda _{\left( \alpha \right) k}^{i}}\lambda
_{\left( \alpha \right) }^{l}\frac{\partial \delta x^{i}}{\partial x^{l}}]
\notag \\
&&+\left[ L_{\lambda }\right] _{\lambda _{\left( \alpha \right) }^{i}}\frac{%
\partial \delta x^{i}}{\partial x^{l}}\lambda _{\left( \alpha \right) }^{l}-%
\left[ L_{\lambda }\right] _{\lambda _{\left( \alpha \right) }^{i}}\lambda
_{\left( \alpha \right) l}^{i}\delta x^{l}
=0.
\end{eqnarray}%
With the relation (see Appedix (I.1))%
\begin{equation}
\frac{\partial }{\partial x^{k}}\left\{ \left[ L_{\lambda }\right] _{\lambda
_{\left( \alpha \right) }^{l}}\lambda _{\left( \alpha \right) }^{k}\right\}
=-\left[ L_{\lambda }\right] _{\lambda _{\left( \alpha \right) }^{i}}\lambda
_{\left( \alpha \right) l}^{i}
\end{equation}%
one can obtain%
\begin{equation}
\left[ L_{\lambda }\right] _{\lambda _{\left( \alpha \right) }^{i}}\lambda
_{\left( \alpha \right) l}^{i}\delta x^{l}=-\frac{\partial }{\partial x^{k}}%
\left\{ \left[ L_{\lambda }\right] _{\lambda _{\left( \alpha \right)
}^{l}}\lambda _{\left( \alpha \right) }^{k}\delta x^{l}\right\} +\left[
L_{\lambda }\right] _{\lambda _{\left( \alpha \right) }^{l}}\lambda _{\left(
\alpha \right) }^{k}\frac{\partial \delta x^{l}}{\partial x^{k}}.
\end{equation}%
By (51), eq. (49) can be simplified as%
\begin{equation}
\frac{\partial }{\partial x^{k}}\left\{ \left[ \left( L_{\lambda }\delta
_{l}^{k}-\frac{\partial L_{\lambda }}{\partial \lambda _{\left( \alpha
\right) k}^{i}}\lambda _{\left( \alpha \right) l}^{i}\right) +\left[
L_{\lambda }\right] _{\lambda _{\left( \alpha \right) }^{l}}\lambda _{\left(
\alpha \right) }^{k}\right] \delta x^{l}+\frac{\partial L_{\lambda }}{%
\partial \lambda _{\left( \alpha \right) k}^{i}}\lambda _{\left( \alpha
\right) }^{l}\frac{\partial \delta x^{i}}{\partial x^{l}}\right\} =0.
\end{equation}%
which is again the conservation law corresponding to transformation (44) and
(45). However, it is now a conservation law given by gravitational Lagrangian
only,  different from (48).

Up to now, we studied the general conservation law. Next, we will consider
the energy-momentum conservation law. The generally covariant conservation
law for the energy-momentum corresponds to the general translation
transformation. Let $\delta x^{i}$ in Eq. (44) is the vector in the
Riemannian manifold,%
\begin{equation}
\delta x^{l}=\lambda _{\left( \beta \right) }^{l}\delta x_{\left( \beta
\right) }
\end{equation}%
the general translation transformation is%
\begin{equation*}
\delta x_{\left( \beta \right) }=a_{\left( \beta \right) },\ \ \ \ \left(
\beta =1,2,3,4\right)
\end{equation*}%
where $a_{\left( \beta \right) }$ are infinitesimal parameters for
translation independent of $x$. When $\lambda _{\left( \beta \right) }$ is
fixed, the translation transformation is unique. Therefore, the general
translation transformation can be expressed as%
\begin{equation}
x^{l\prime }=x^{l}+\lambda _{\left( \beta \right) }^{l}a_{\left( \beta
\right) }.
\end{equation}%
Without gravitational field, $\lambda _{\left( \beta \right) }^{l}=\delta
_{\left( \beta \right) }^{l}$, (54) reduces to the trivial translation
tranformation $x^{l\prime }=x^{l}+a^{l}$. With%
\begin{equation}
\frac{\partial \delta x^{i}}{\partial x^{l}}=\frac{\partial \lambda _{\left(
\alpha \right) }^{i}}{\partial x^{l}}a_{\left( \alpha \right) }
\end{equation}%
and substituting (53) to (52), we find the conservation law corresponding to
the general translation transformation%
\begin{equation}
\frac{\partial }{\partial x^{k}}\left\{ \left( L_{\lambda }\delta _{l}^{k}-%
\frac{\partial L_{\lambda }}{\partial \lambda _{\left( \alpha \right) k}^{i}}%
\lambda _{\left( \alpha \right) l}^{i}\right) \lambda _{\left( \beta \right)
}^{l}+\left[ L_{\lambda }\right] _{\lambda _{\left( \alpha \right)
}^{l}}\lambda _{\left( \alpha \right) }^{k}\lambda _{\left( \beta \right)
}^{l}+\frac{\partial L_{\lambda }}{\partial \lambda _{\left( \alpha \right)
k}^{i}}\lambda _{\left( \alpha \right) }^{l}\frac{\partial \lambda _{\left(
\beta \right) }^{l}}{\partial x^{l}}\right\} =0.
\end{equation}%
With (37) and (31), the second term in the brace bracket of (56) has direct
relation to energy-momentum tensor of matter $T_{\left( \beta \right) }^{k}
$%
\begin{equation}
T_{\left( \beta \right) }^{k}=\frac{1}{\sqrt{g}}\left[ L_{\lambda }\right]
_{\lambda _{\left( \alpha \right) }^{l}}\lambda _{\left( \alpha \right)
}^{k}\lambda _{\left( \alpha \right) }^{l}.
\end{equation}%
Defining%
\begin{equation}
t_{\left( \beta \right) }^{k}=\frac{1}{\sqrt{g}}\left[ \left( L_{\lambda
}\delta _{l}^{k}-\frac{\partial L_{\lambda }}{\partial \lambda _{\left(
\alpha \right) k}^{i}}\lambda _{\left( \alpha \right) l}^{i}\right) \lambda
_{\left( \beta \right) }^{l}+\frac{\partial L_{\lambda }}{\partial \lambda
_{\left( \alpha \right) k}^{i}}\lambda _{\left( \alpha \right) }^{l}\frac{%
\partial \lambda _{\left( \beta \right) }^{l}}{\partial x^{l}}\right]
\end{equation}%
Eq. (56) can be simplifed as%
\begin{equation}
\frac{\partial }{\partial x^{k}}\left[ \sqrt{g}\left( T_{\left( \beta
\right) }^{k}+t_{\left( \beta \right) }^{k}\right) \right] =0.
\end{equation}%
Since $T_{\left( \beta \right) }^{k}$ is the energy-momentum tensor of
matter, $t_{\left( \beta \right) }^{k}$ defined in (58) should be the
energy-momentum tensor for the gravitational field. These tensors have both
semi-metric index and Riemannian index, which is essentially different from
those energy-momentum tensor with only Riemannian index.

When the system is closed, i.e. the total energy-momentum tensor vanishes at
the infinity of 3-dimensional space (see Appendix III), using Eq. (59) and
4-dimensional Gaussian theorem we obtain%
\begin{equation}
\int_{\sigma _{1}}\left( T_{\left( \beta \right) }^{k}+t_{\left( \beta
\right) }^{k}\right) \sqrt{g}d\sigma _{k}=\int_{\sigma _{2}}\left( T_{\left(
\beta \right) }^{k}+t_{\left( \beta \right) }^{k}\right) \sqrt{g}d\sigma
_{k}=const
\end{equation}%
where 4-dimensional spacetime is made by the hypersurface $\sigma _{1}$ and $%
\sigma _{2},$ and the infinity side face $\Sigma $. It is clear that%
\begin{equation}
P_{\left( \alpha \right) }=\frac{1}{ic}\int_{\sigma }\left( T_{\left( \alpha
\right) }^{k}+t_{\left( \alpha \right) }^{k}\right) \sqrt{g}d\sigma _{k}
\end{equation}%
is the conserved quantity. This conserved quantity is obtained under the
general translation transformation (54), therefore it is the four momentum
of gravity and matter. When $\sigma $ is chosen as the hypersurface
perpendicular to time axes $t$, (61) can be written as the 3-dimensional
volume intergral%
\begin{equation}
P_{\left( \alpha \right) }=\frac{1}{ic}\int_{V}\left( T_{\left( \alpha
\right) }^{4}+t_{\left( \alpha \right) }^{4}\right) \sqrt{g}dV.
\end{equation}%
Then%
\begin{equation}
G_{\left( \alpha \right) }=\frac{1}{ic}\left( T_{\left( \alpha \right)
}^{4}+t_{\left( \alpha \right) }^{4}\right) \sqrt{g}
\end{equation}%
is the density of total 4-momentum which leads to%
\begin{equation}
P_{\left( \alpha \right) }=\int G_{\left( \alpha \right) }dV.
\end{equation}%
Here one should pay attention to the index $\left( \alpha \right) $ of
4-momentum $P_{\left( \alpha \right) }$. It is the index of semi-metric and is
different from the ordinary theory. \ Einstein and Moller used the covariant
index to define their 4-momentum, Landau used the contrariant index. Later
we will discuss the physical meaning of semi-metric index and its advantages.

Next we will study the specific expressions of $t_{\left( \alpha \right) }^{k}$ and $%
T_{\left( \alpha \right) }^{k}+t_{\left( \alpha \right) }^{k}$. With the
antisymmetric relation between $k$ and $l$ (see appendix (II.4))%
\begin{equation}
\frac{\partial L_{\lambda }}{\partial \lambda _{\left( \alpha \right) k}^{k}}%
\lambda _{\left( \alpha \right) }^{l}=-\frac{\partial L_{\lambda }}{\partial
\lambda _{\left( \alpha \right) l}^{i}}\lambda _{\left( \alpha \right) }^{k},
\end{equation}%
$t_{\left( \alpha \right) }^{k}$ defined in (58) can be rewritten as%
\begin{equation}
t_{\left( \alpha \right) }^{k}=\frac{1}{\sqrt{g}}\left\{ L_{\lambda }\delta
_{l}^{k}\lambda _{\left( \beta \right) }^{l}-\left[ \frac{\partial
L_{\lambda }}{\partial \lambda _{\left( \alpha \right) k}^{i}}\lambda
_{\left( \alpha \right) l}^{i}\lambda _{\left( \beta \right) }^{l}+\frac{%
\partial L_{\lambda }}{\partial \lambda _{\left( \alpha \right) l}^{i}}%
\lambda _{\left( \beta \right) l}^{i}\lambda _{\left( \alpha \right) }^{k}%
\right] \right\} .
\end{equation}%
Substituting the expression of $L_{\lambda }$ (26) into (66)and using relation
(27) and (28),  we obtain the expression for the
energy-momentum tensor $t_{\left( \alpha \right) }^{k}$ for gravitational
field after lengthy calculation%
\begin{eqnarray}
t_{\left( \alpha \right) }^{k} &=&\frac{c^{4}}{16\pi k}\{\lambda _{\left(
\alpha \right) }^{k}\left[ \eta _{\left( \beta \right) }\eta _{\left( \beta
\right) }-\eta _{\left( \delta \beta \gamma \right) }\eta _{\left( \gamma
\beta \delta \right) }\right] -2\lambda _{\left( \beta \right) }\left[ \eta
_{\left( \beta \right) }\eta _{\left( \alpha \right) }-\eta _{\left( \delta
\beta \gamma \right) }\eta _{\left( \gamma \alpha \delta \right) }\right]
\notag \\
&&-2\lambda _{\left( \gamma \right) }^{k}\left[ \eta _{\left( \beta \right)
}\eta _{\left( \alpha \beta \gamma \right) }+\eta _{\left( \beta \right)
}\eta _{\left( \beta \alpha \gamma \right) }\right] +2\lambda _{\left(
\delta \right) }^{k}\eta _{\left( \alpha \beta \gamma \right) }\eta _{\left(
\gamma \beta \delta \right) }
\end{eqnarray}%
which is very useful when we calculate the energy of gravitional field.

With the above expression and (61) one can find that the 4-momentum of
gravitational field defined by $t_{\left( \alpha \right) }^{k}$ is
determined totally by Ricci coefficient $\eta _{\left( \alpha \beta \gamma
\right) }$ and $\eta _{\left( \alpha \right) }$ due to the contraction
between index $k$ in $\lambda _{\left( \alpha \right) }^{k}$ and $\sqrt{g}%
d\sigma _{k}$. From Ref. [12] and [13],  the existence of gravitation field is
fully determined by nonvanishing Ricc coefficients, which confirms that $%
t_{\left( \alpha \right) }^{k}$ is indeed the energy-momentum tensor of
gravitational field and there is no inertial part in the 4-momentum defined
by $t_{\left( \alpha \right) }^{k}$. In vacuum, i.e. the flat spacetime all
Ricci coefficient $\eta _{\left( \alpha \beta \beta \right) }$ and $\eta
_{\left( \alpha \right) }$ are zero in any coordinates. Therefore the
energy-momentum tensor $t_{\left( \alpha \right) }^{k}$ and its 4-momentum
are zero in arbitrary coordinates, which solves the puzzle appears in those
theories proposed by Einstein et al.

Furthermore, with the relation from Appendix (II.3)%
\begin{equation}
\left\{ \left( L_{\lambda }\delta _{l}^{k}-\frac{\partial L_{\lambda }}{%
\partial \lambda _{\left( \alpha \right) k}^{i}}\lambda _{\left( \alpha
\right) l}^{i}\right) +\left[ L_{\lambda }\right] _{\lambda _{\left( \alpha
\right) }^{l}}\lambda _{\left( \alpha \right) }^{k}\right\} =-\frac{\partial
}{\partial x^{j}}\left[ \frac{\partial L_{\lambda }}{\partial \lambda
_{\left( \alpha \right) j}^{l}}\lambda _{\left( \alpha \right) }^{k}\right]
\end{equation}%
one can find an important result%
\begin{eqnarray}
&&\left[ L_{\lambda }\delta _{l}^{k}\lambda _{\left( \beta \right)
}^{l}-\left( \frac{\partial L_{\lambda }}{\partial \lambda _{\left( \alpha
\right) k}^{i}}\lambda _{\left( \alpha \right) l}^{i}\lambda _{\left( \beta
\right) }^{l}+\frac{\partial L_{\lambda }}{\partial \lambda _{\left( \alpha
\right) l}^{i}}\lambda _{\left( \beta \right) l}^{i}\lambda _{\left( \alpha
\right) }^{k}\right) \right] +\left[ L_{\lambda }\right] _{\lambda _{\left(
\alpha \right) }^{l}}\lambda _{\left( \alpha \right) }^{k}\lambda _{\left(
\beta \right) }^{l}  \notag \\
&=&-\frac{\partial }{\partial x^{j}}\left[ \frac{\partial L_{\lambda }}{%
\partial \lambda _{\left( \alpha \right) j}^{l}}\lambda _{\left( \alpha
\right) }^{k}\lambda _{\left( \beta \right) }^{l}\right] .
\end{eqnarray}%
Substituting (57) and (66) into above expression and utilizing (65), the total
energy-momentum tensor $T_{\left( \beta \right) }^{k}+t_{\left( \beta
\right) }^{k}$ can be simplified as the following 4-dimensional divergence%
\begin{equation}
\sqrt{g}\left( T_{\left( \beta \right) }^{k}+t_{\left( \beta \right)
}^{k}\right) =\frac{\partial v_{\left( \beta \right) }^{kj}}{\partial x^{j}},
\end{equation}%
where%
\begin{equation}
v_{\left( \beta \right) }^{kj}=\frac{1}{2}\left[ \lambda _{\left( \beta
\right) }^{i}\left( \lambda _{\left( \alpha \right) }^{j}\frac{\partial
L_{\lambda }}{\partial \lambda _{\left( \alpha \right) k}^{i}}-\lambda
_{\left( \alpha \right) }^{k}\frac{\partial L_{\lambda }}{\partial \lambda
_{\left( \alpha \right) j}^{i}}\right) \right] ,
\end{equation}%
and the index $k$ and $j$ are antisymmetric. Using the expression of $%
L_{\lambda }$ in (24), after lengthy calculation one can find the expression
of $v_{\left( \alpha \right) }^{kj}$%
\begin{equation}
v_{\left( \alpha \right) }^{kj}=\sqrt{g}V_{\left( \alpha \right) }^{kj},
\end{equation}%
\begin{equation}
V_{\left( \alpha \right) }^{kj}=\frac{c^{4}}{8\pi k}\left[ \lambda _{\left(
\alpha \right) }^{i}\lambda _{\left( \beta \right) }^{k}\left( \lambda
_{\left( \beta \right) }^{j}\right) _{i}+\left( \lambda _{\left( \alpha
\right) }^{k}\lambda _{\left( \beta \right) }^{j}-\lambda _{\left( \alpha
\right) }^{j}\lambda _{\left( \beta \right) }^{k}\right) \left( \lambda
_{\left( \beta \right) }^{i}\right) _{i}\right] ,
\end{equation}%
or expressed by the Ricci coefficients%
\begin{equation}
V_{\left( \alpha \right) }^{kj}=\frac{c^{4}}{8\pi k}\left[ \lambda _{\left(
\beta \right) }^{k}\lambda _{\left( \gamma \right) }^{j}\eta _{\left( \alpha
\beta \gamma \right) }+\left( \lambda _{\left( \alpha \right) }^{k}\lambda
_{\left( \beta \right) }^{j}-\lambda _{\left( \alpha \right) }^{j}\lambda
_{\left( \beta \right) }^{k}\right) \eta _{\left( \beta \right) }\right] .
\end{equation}%
In terms of the antisymmetric property of $k$ and $j$ in $v_{\left( \alpha
\right) }^{kj}$, with (70), (62) and 3-dimensional Gaussian theorem, the
4-momentum $P_{\left( \alpha \right) }$ can be expressed as a surface
integral%
\begin{equation}
P_{\left( \alpha \right) }=\frac{1}{ic}\int_{S}v_{\left( \alpha \right)
}^{4j}dS_{j},
\end{equation}%
where $S$ is the closed surface to enclose the 3-dimensional volume $V$.
When we consider the closed system, $S$ is the close surface at infinity.
Thus $P_{\left( \alpha \right) }$ is only determined by the quantities on
the surface at infinity, which will be very convenient for the actual
calculations.

At the end of this section, we discuss the general covariance of our energy-momentum
conservation law. Firstly from (73) it is evident that $V_{\left( \alpha
\right) }^{kj}$ is a 2-rank Riemannian tensor with antisymmetric index $k$
and $j$ expressed in terms of semi-metric $\lambda _{\left( \alpha \right) }^{i}$ and
its covariant derivatives ( index $\left( \alpha \right)
$ is vectorial in the semi-metric representation and a scalar in Riemannian space
[9,12,13]). Then with (70) and (72) the total energy-momentum tensor is the
covariant derivative of $V_{\left( \alpha \right) }^{kj}$%
\begin{equation}
T_{\left( \alpha \right) }^{k}+t_{\left( \alpha \right) }^{k}=\frac{1}{\sqrt{%
g}}\frac{\partial \left( \sqrt{g}V_{\left( \alpha \right) }^{kj}\right) }{%
\partial x^{j}}=\left( V_{\left( \alpha \right) }^{kj}\right) _{j}.
\end{equation}%
With this result it is clear that $T_{\left( \alpha \right) }^{k}+t_{\left(
\alpha \right) }^{k}$ is a Riemannian tensor for index $k$. Since $T_{\left(
\alpha \right) }^{k}=T_{i}^{k}\lambda _{\left( \alpha \right) }^{i}$ and $%
T_{i}^{k}$ is a $(1+1)$ tensor in the Riemannian space,  $T_{\left(
\alpha \right) }^{k}$ is also a Riemannian tensor for index $k$. Hence, the gravitational energy-momentum tensor $t_{\left( \alpha \right) }^{k}$ is also a Riemann tensor with respect to index $k$, as can be seen from (67).

Since $T_{\left( \alpha \right) }^{k}+t_{\left( \alpha \right) }^{k}$ is a
Riemannian tensor, the conservation law (59) of gravitational and matter
field can be written as the generally covariant divergence%
\begin{equation}
\frac{1}{\sqrt{g}}\frac{\partial \left[ \sqrt{g}\left( T_{\left( \alpha
\right) }^{k}+t_{\left( \alpha \right) }^{k}\right) \right] }{\partial x^{k}}%
=\left( T_{\left( \alpha \right) }^{k}+t_{\left( \alpha \right) }^{k}\right)
_{k}=0.
\end{equation}%
Substituting (76) to (61) one can obtain the 4-momentum%
\begin{equation}
P_{\left( \alpha \right) }=\frac{1}{ic}\int_{\sigma }\left( T_{\left( \alpha
\right) }^{k}+t_{\left( \alpha \right) }^{k}\right) \sqrt{g}d\sigma _{k}=%
\frac{1}{ic}\int_{\sigma }\left( V_{\left( \alpha \right) }^{kj}\right) _{j}%
\sqrt{g}d\sigma _{k}.
\end{equation}%
Since $\sqrt{g}d\sigma _{k}$ is a covariant surface element in the
4-dimensional Riemann manifold, the above formular of $P_{\left( \alpha \right) }
$ is valid for arbitrary coodinates. Furthermore we can express (78) as a
2-dimensional integral
\begin{equation}
P_{\left( \alpha \right) }=\frac{1}{ic}\int_{S}V_{\left( \alpha \right)
}^{kj}\sqrt{g}dS_{kj}=\frac{1}{ic}\int_{S}v_{\left( \alpha \right)
}^{kj}dS_{kj}.
\end{equation}%
Eq. (75) is the special case of (79).

From (79) or (78) one can conclude that $P_{\left( \alpha \right) }$ is a
vector with semi-metric index, a scalar in the Riemannian manifold,
meaning that it is invariant for general transformation group $\left( a\right)
$ and covariant for orthogonal transformation group $\left( \sigma \right) $%
. With appendix IV we prove that under general circumstances, $\lambda _{i\left( \alpha \right) }$ can be
either uniquely determined by Einstein equation and coordinate conditions or related to each other by
a orthogonal transformation $L_{\left( \alpha \beta \right) }$.
From (79), (73) and (72), $P_{\left( \alpha \right) }$ corresponding to two sets of $\lambda
_{i\left( \alpha \right) }$ are related by%
\begin{equation}
P_{\left( \alpha \right) }^{\prime }=L_{\left( \alpha \beta \right)
}P_{\left( \beta \right) }
\end{equation}%
where $L_{\left( \alpha \beta \right) }$ is an orthogonal matrix independent
of $x$. For a closed system, $P_{\left( \alpha \right) }$ is only determined
by the value of $v_{\left( \alpha \right) }^{kj}$ on the hypersurface $S$ at
infinity. With (13), (14) and discussion in Sec. II, we know that the
orthogonal transformation (80) independent of $x$ is just the Lorentz
transformation. Since $P_{\left( \alpha \right) }$ is a conserved quantity
for a closed system, the system can move inertially only. It is well-known
that inertial systems are related by Lorentz transformation, which is the
physical meanning of 4-momentum expressed by semi-metric index and
transformation (80).

From above discussions one can find that energy-momentum tensor (76),
energy-momentum conservation law (77), and 4-momentum (78), (79) are
strictly generally covariant for arbitrary coordinates in the Riemannian
manifold. Therefore, for the total energy of closed systems we will obtain
the reasonable results in either  spherical coordinates or any other
non-quasi-Galilean coordinates. Moreover, since the 4-momentum
related to $t_{\left( \alpha \right) }^{i}$ includes only gravitational
field without the inertial part, $t_{\left( \alpha \right) }^{i}$ and
the corresponding 4-momentum  always vanish in the vacuum without
matter and gravitational field in arbitrary coordinates. At last, $%
t_{\left( \alpha \right) }^{i}$ decays into zero by $1/r^{4}$ at infinity which
guarantees the existence of conserved quantities. Therefore, our theory overcomes the difficulties
 of conservation laws proposed by Einstein, Moller
and Landau  discussed in the introduction.

\section{A simple example}

Finally we check our theory by a simple example. The metric with spherical
distribution of matter can be solved strictly from Einstein equation (3).
Let's calculate the total energy of such a system. In this case due to $%
g_{ik}=0$ with $i\neq k$, let's denote%
\begin{equation}
g_{ii}=H_{i}^{2},\ \ \ \ g^{ii}=\frac{1}{H_{i}^{2}},\ \ \ \ \sqrt{g}%
=H_{1}H_{2}H_{3}H_{4},
\end{equation}%
then we have with (8)%
\begin{equation}
\lambda _{i\left( i\right) }=H_{i},\ \ \ \lambda _{\left( i\right) }^{i}=%
\frac{1}{H_{i}},\ \ \ \lambda _{\left( \alpha \right) }^{i}=\lambda
_{i\left( \alpha \right) }=0,\ \ at\ i\neq a,
\end{equation}%
\begin{equation}
-dS^{2}=g_{ik}dx^{i}dx^{k}=H_{i}^{2}\left( dx^{i}\right) ^{2}.
\end{equation}%
Substituting (82) to (27) and (28), one obtains the Ricci coefficients%
\begin{equation}
\eta _{\left( \alpha \beta \gamma \right) }=\frac{1}{4H_{\alpha }H_{\beta
}H_{\gamma }}\left\{ \frac{\partial }{\partial x^{\beta }}\left[ H_{\alpha
}^{2}+H_{\gamma }^{2}\right] \delta _{\gamma }^{\alpha }-\frac{\partial }{%
\partial x^{\gamma }}\left[ H_{\alpha }^{2}+H_{\beta }^{2}\right] \delta
_{\beta }^{\alpha }\right\}
\end{equation}%
where the repeated indices don't sum (From now on, all summation will be
indicated explicitly)%
\begin{equation}
\eta _{\left( \alpha \right) }=\frac{1}{2}\sum_{i=1,\ i\neq a}^{4}\frac{1}{%
H_{\alpha }H_{i}^{2}}\frac{\partial H_{i}^{2}}{\partial x^{a}}.
\end{equation}%
From (75) $P_{\left( \alpha \right) }$ is only related to the value of $%
v_{\left( \alpha \right) }^{kj}$ at infinity, i.e. only related to $H_{i}$
at infinity. For a spherical symmtric distribution of matter, $g_{ik}$ or $%
H_{i}$ can be obtained at large distance $r$%
\begin{equation}
-dS^{2}=\left( 1+\frac{2kM}{c^{2}r}\right) \left(
dx_{1}^{2}+dx_{2}^{2}+dx_{3}^{3}\right) +\left( 1-\frac{2kM}{c^{2}r}\right)
dx_{4}^{2},
\end{equation}%
\begin{equation*}
x_{4}=ict,\ \ \ r^{2}=x_{1}^{2}+x_{2}^{2}+x_{3}^{2},
\end{equation*}%
where $M$ is the total mass of the system. Then we find%
\begin{equation}
H_{1}^{2}=H_{2}^{2}=H_{3}^{2}=1+\frac{2kM}{c^{2}r},\ \ \ H_{4}^{2}=1-\frac{%
2kM}{c^{2}r}.
\end{equation}%
Substituting (87) to (84) and (85) and using (72) and (74) one obtains%
\begin{equation*}
v_{\left( 4\right) }^{41}=-\frac{c^{2}M}{4\pi }\frac{x_{1}}{H_{1}r^{3}},
\end{equation*}%
\begin{equation*}
v_{\left( 4\right) }^{42}=-\frac{c^{2}M}{4\pi }\frac{x_{2}}{H_{1}r^{3}},
\end{equation*}%
\begin{equation*}
v_{\left( 4\right) }^{43}=-\frac{c^{2}M}{4\pi }\frac{x_{3}}{H_{1}r^{3}},
\end{equation*}%
\begin{equation}
v_{\left( a\right) }^{4j}=0,\ \ \ \left( \alpha \neq 4\right) .
\end{equation}%
With (75) we can calculate%
\begin{eqnarray*}
P_{\left( 4\right) } &=&\frac{1}{ic}\int_{S}v_{\left( 4\right) }^{4j}dS_{j}=%
\frac{1}{ic}\int_{S}v_{\left( 4\right) }^{4j}n_{j}dS \\
&=&\frac{icM}{4\pi }\int \frac{1}{H_{1}}\frac{x_{j}^{2}}{r^{4}}d\Omega
|_{r\rightarrow \infty }=iMc\frac{1}{H_{1}}|_{r\rightarrow \infty }=iMc,
\end{eqnarray*}%
\begin{equation*}
d\Omega =r^{2}\sin \theta d\theta d\phi ,\ \ \ n_{i}=\frac{x^{i}}{r},
\end{equation*}%
\begin{equation}
P_{\left( a\right) }=0,\ \ \ \ \ when\ a\neq 4.
\end{equation}%
i.e%
\begin{equation}
P_{\left( \alpha \right) }=i\delta _{\left( \alpha 4\right) }Mc.
\end{equation}%
Since
\begin{equation*}
P_{\left( 4\right) }=i\frac{E}{c},
\end{equation*}%
we obtain%
\begin{equation*}
E=Mc^{2},
\end{equation*}
which is correct answer for the system.

Based on our theory, we also calculated many-body problems and the radiation
of gravitational field and obtained the reasonable results, which will be
presented elsewhere.

\appendix
\renewcommand{\thesection}{\Roman{section}}

\section{}

Due to
\begin{equation*}
\left( R_{i}^{k}-\frac{1}{2}\delta _{i}^{k}R\right) _{k}=\frac{8\pi k}{c^{4}}%
\left( T_{i}^{k}\right) _{k}=0,
\end{equation*}%
we have
\begin{equation*}
\left( T_{i}^{k}\right) _{k}=\frac{1}{\sqrt{g}}\frac{\partial \left( \sqrt{g}%
T_{i}^{k}\right) }{\partial x^{k}}-\frac{1}{2}\frac{\partial g_{kl}}{%
\partial x^{i}}T^{kl}=0.
\end{equation*}%
With (8) and (9) and $g_{ki}g^{kl}=\delta _{i}^{l}$, the above expression
can be rewritten as%
\begin{equation*}
\frac{\partial \left( \sqrt{g}T_{i}^{k}\right) }{\partial x^{k}}=\frac{1}{2}%
\sqrt{g}\frac{\partial g_{kl}}{\partial x^{i}}T^{kl}=-\sqrt{g}\frac{\partial
\lambda _{\left( \alpha \right) }^{l}}{\partial x^{i}}T_{l\left( \alpha
\right) }.
\end{equation*}%
Substituting (37) to above equation, with (33) we find%
\begin{equation}
\frac{\partial }{\partial x^{k}}\left( \left[ L_{\lambda }\right] _{\lambda
_{\left( \alpha \right) }^{i}}\lambda _{\left( \alpha \right) }^{k}\right) =-%
\left[ L_{\lambda }\right] _{\lambda _{\left( \alpha \right) }^{l}}\lambda
_{\left( \alpha \right) i}^{l}.
\end{equation}

\section{}

Since $L_{\lambda }$ is generally covariant, i.e. it is  invariant under
general transformation $\left( a\right) $, it is also invariant for
translation%
\begin{equation*}
x^{i\prime }=x^{i}+a^{i},\ \ \ \ i=1,2,3,4,
\end{equation*}%
where $a^{i}$ is infinitesimal translation parameters independent of $x$, i.e%
\begin{equation*}
\delta x^{i}=a^{i},\ \ \ \frac{\partial \delta x^{i}}{\partial x^{l}}=0.
\end{equation*}%
With (52) we find%
\begin{equation}
\frac{\partial }{\partial x^{k}}\left\{ \left( L_{\lambda }\delta _{l}^{k}-%
\frac{\partial L_{\lambda }}{\partial \lambda _{\left( \alpha \right) k}^{i}}%
\lambda _{\left( \alpha \right) l}^{i}\right) +\left[ L_{\lambda }\right]
_{\lambda _{\left( \alpha \right) }^{i}}\lambda _{\left( \alpha \right)
}^{k}\right\} =0.
\end{equation}%
which is the Einstein-Tolman's conservation law in the semi-metric
representation.

From (II.1) and (52) we have an important relation%
\begin{equation}
\left\{ \left( L_{\lambda }\delta _{l}^{k}-\frac{\partial L_{\lambda }}{%
\partial \lambda _{\left( \alpha \right) k}^{i}}\lambda _{\left( \alpha
\right) l}^{i}\right) +\left[ L_{\lambda }\right] _{\lambda _{\left( \alpha
\right) }^{l}}\lambda _{\left( \alpha \right) }^{k}\right\} \frac{\partial
\delta x^{l}}{\partial x^{k}}=-\frac{\partial }{\partial x^{j}}\left[ \frac{%
\partial L_{\lambda }}{\partial \lambda _{\left( \alpha \right) j}^{i}}%
\lambda _{\left( \alpha \right) }^{l}\frac{\partial \delta x^{i}}{\partial
x^{l}}\right] .
\end{equation}

Since $L_{\lambda }$ is generally covariant, it is certainly invariant under the
following infinitesimal orthogonal transformation%
\begin{equation*}
x^{l\prime }=x^{l}+\alpha _{i}^{l}x^{i},
\end{equation*}%
\begin{equation*}
\delta x^{l}=\alpha _{i}^{l}x^{i},\ \ \ \frac{\partial \delta x^{l}}{%
\partial x^{i}}=\alpha _{i}^{l}
\end{equation*}%
where $\alpha _{i}^{l}$ is not a function of $x$. With this tranformation we
obtain%
\begin{equation}
\left\{ \left( L_{\lambda }\delta _{l}^{k}-\frac{\partial L_{\lambda }}{%
\partial \lambda _{\left( \alpha \right) k}^{i}}\lambda _{\left( \alpha
\right) l}^{i}\right) +\left[ L_{\lambda }\right] _{\lambda _{\left( \alpha
\right) }^{l}}\lambda _{\left( \alpha \right) }^{k}\right\} =-\frac{\partial
}{\partial x^{j}}\left[ \frac{\partial L_{\lambda }}{\partial \lambda
_{\left( \alpha \right) j}^{l}}\lambda _{\left( \alpha \right) }^{k}\right] .
\end{equation}%
Substituting (II.3) to (II.1), we have%
\begin{equation*}
\frac{\partial ^{2}}{\partial x^{j}\partial x^{k}}\left[ \frac{\partial
L_{\lambda }}{\partial \lambda _{\left( \alpha \right) j}^{l}}\lambda
_{\left( \alpha \right) }^{k}\right] =0,
\end{equation*}%
which shows that $\frac{\partial L_{\lambda }}{\partial \lambda _{\left(
\alpha \right) j}^{l}}\lambda _{\left( \alpha \right) }^{k}$ is
antisymmetric in indices $j$ and $k$, i.e.%
\begin{equation}
\frac{\partial L_{\lambda }}{\partial \lambda _{\left( \alpha \right) j}^{l}}%
\lambda _{\left( \alpha \right) }^{k}=-\frac{\partial L_{\lambda }}{\partial
\lambda _{\left( \alpha \right) k}^{l}}\lambda _{\left( \alpha \right) }^{j}.
\end{equation}

\section{}

In order for (60) to be valid, $T_{\left( \beta \right) }^{i}+t_{\left( \beta
\right) }^{i}$ must decay into zero at infinity faster than $1/r^{3}$ (see
ref. [10]). Since $T_{\left( \beta \right) }^{k}=\lambda _{\left( \beta
\right) }^{i}T_{i}^{k}$ is always zero at infinity far from the matter, it
is enough to ask $t_{\left( \beta \right) }^{i}$ to satisfy above
requirement. From (67) $t_{\left( \beta \right) }^{i}$ is a product of Ricci
coefficients containing only the first derivative of $%
\lambda _{i\left( \alpha \right) }$ which is proportional to $1/r^{2}$ at
infinity. Therefore, $t_{\left( \beta \right) }^{i}$ is proportional to $%
1/r^{4}$ at infinity which meets the criteria of (60).

\section{Proof of uniqueness of $\protect\lambda _{i\left(
\protect\alpha \right) }$}

Since $||g_{ik}||$ is a symmetric matrix, and all $g\binom{1...k}{1...k}\neq
0$ $(k=1,2,3,4)$, $||g_{ik}||$ can be always decomposed as a product of
lower triangle matrix and its transpose [19]%
\begin{equation*}
||g_{ik}||=||\lambda _{i\left( \alpha \right) }||\ ||\lambda _{k\left(
\alpha \right) }||^{T},
\end{equation*}%
or%
\begin{equation}
g_{ik}=\lambda _{i\left( \alpha \right) }\lambda _{\left( \alpha \right)
k}^{T}=\lambda _{i\left( \alpha \right) }\lambda _{k\left( \alpha \right) },
\end{equation}%
where%
\begin{equation}
||\lambda _{i\left( \alpha \right) }||=\left(
\begin{array}{cccc}
\lambda _{1\left( 1\right) } & 0 & 0 & 0 \\
\lambda _{2\left( 1\right) } & \lambda _{2\left( 2\right) } & 0 & 0 \\
\lambda _{3\left( 1\right) } & \lambda _{3\left( 2\right) } & \lambda
_{3\left( 3\right) } & 0 \\
\lambda _{4\left( 1\right) } & \lambda _{4\left( 2\right) } & \lambda
_{4\left( 3\right) } & \lambda _{4\left( 4\right) }%
\end{array}%
\right) .
\end{equation}%
which shows that there are only 10 nonzero components for $||\lambda
_{i\left( \alpha \right) }||$. When we obtain ten components of $g_{ik}$ by
solving Einstein equation and coordinate conditions,  we can
uniquely find ten components of $\lambda _{i\left( \alpha \right) }$ with (IV.1).

On the other hand, suppose any matrix $||\lambda _{i\left( \alpha \right) }||$
satisfies%
\begin{equation*}
g_{ik}=\lambda _{i\left( \alpha \right) }\lambda _{k\left( \beta \right) }.
\end{equation*}
Since (IV.1) keeps invariant under orthogonal transformation group $\left(
\sigma \right) $ which is determined by six parameters, we can always choose
proper parameters to make $\lambda _{i\left( \alpha \right) }$  a lower
triangular matrix [11]. Thefore, any matrix $\lambda _{i\left( \alpha
\right) }$ satisfying (IV.1) can be related to a lower triangular matrix via
orthogonal tranformation.

\

\end{document}